\documentclass[11pt,a4paper]{article}
\usepackage{jheppub}
\usepackage[utf8]{inputenc}
\usepackage{graphicx,fancybox,float,comment,bm}
\usepackage{tensor}

\usepackage[dvipsnames]{xcolor}
\usepackage{amsmath,amssymb,amsfonts}
\usepackage[bbgreekl]{mathbbol}
\usepackage{soul}
\usepackage[symbol]{footmisc}

\usepackage{ulem}

\title{A supersymmetric spin current}
\author[1]{Casey Cartwright, }
\author[2]{Domingo Gallegos, }
\author[1]{Umut G\"ursoy,\footnote[3]{Deceased} }
\author[3]{Roi Klein, }
\author[3]{Amos Yarom }
\date{September 2022}
\dedicated{To Umut---for his insight, intellect, and friendship.}

\affiliation[1]{Institute for Theoretical Physics, Utrecht University, Princetonplein 5, 3584 CC Utrecht, The Netherlands}
\affiliation[2]{Facultad de Ciencias, Universidad Nacional Autonoma de Mexico, Investigacion Cientifica, 04510 Ciudad de Mexico, Mexico}
\affiliation[3]{Department of Physics, Technion, Haifa 32000, Israel}

\emailAdd{c.c.cartwright@uu.nl}
\emailAdd{d.gallegos@ciencias.unam.mx}
\emailAdd{u.gursoy@uu.nl}
\emailAdd{roi.klein@campus.technion.ac.il}
\emailAdd{amos.yarom@gmail.com}

\begin{document}
\newcommand{\tet}[2]{e^{#1}_{#2}}
\newcommand{\exd}{\mathrm{d}}
\newcommand{\tr}{\text{tr}}
\newcommand{\ha}{\hat{a}}
\newcommand{\hb}{\hat{b}}
\newcommand{\hc}{\hat{c}}
\newcommand{\hd}{\hat{d}}
\newcommand{\he}{\hat{e}}
\newcommand{\theA}{\mathcal{A}}
\newcommand{\theB}{\mathcal{B}}
\newcommand{\theC}{\mathcal{C}}
\newcommand{\theL}{V}

\newcommand{\AY}[1]{{\textcolor{Red}{ [AY:#1] }}}
\newcommand{\RK}[1]{{\textcolor{Green}{ [RK:#1] }}}
\newcommand{\RKtodo}[1]{{\textcolor{purple}{ [RK (To do list):#1] }}}
\newcommand{\CC}[1]{{\textcolor{Orange}{ [CC:#1] }}}
\newcommand{\UG}[1]{{\textcolor{Blue}{ [UG:#1] }}}

\abstract{We study the supersymmetric structure of the spin current in four dimensional $\mathcal{N}=1$ supersymmetric theories. By coupling the stress tensor multiplet to a vierbein multiplet we identify a spin current with the Hodge dual of the bottom component of the stress tensor multiplet in a wide range of theories.  This implies that in holographic theories the Hodge dual of the $R$ current may serve as a spin current, paving the way for holographic studies of theories with background torsion.}

\maketitle

\section{Introduction}
Conservation of angular momentum follows from invariance of the theory under rotations. In a relativistic theory this conservation law is captured by the existence of a symmetric stress tensor. Indeed, coupling scalar matter fields to a non dynamical background metric guarantees the existence of a manifestly symmetric stress tensor. Similarly, coupling fermionic or vector matter fields to a background vielbein results in a stress tensor which may be made symmetric under the equations of motion. 

Given a symmetric stress tensor, $T_s^{\mu\nu}$, the angular momentum density, $J^{\mu\nu\rho}$, may be expressed entirely in terms of an orbital component which is conserved due to  momentum conservation and symmetry of the stress tensor
\begin{equation}
\label{E:Jsymm}
	J^{\mu\nu\rho} = x^{\rho} T_s^{\mu\nu} - x^{\nu} T_s^{\mu\rho}\,.
\end{equation}
But more generally, the angular momentum density is expressed in terms of an orbital angular momentum component and a spin component which we refer to as the spin current, $S^{\mu\nu\rho}$
\begin{equation}
\label{E:mainJ}
	J^{\mu\nu\rho} = x^{\rho} T^{\mu\nu} - x^{\nu} T^{\mu\rho} + S^{\mu\nu\rho}\,.
\end{equation}
As reviewed in section \ref{S:Torsion} one can go from \eqref{E:Jsymm} to \eqref{E:mainJ} and back by adding improvement terms to the spin current and stress tensor.

Because the stress tensor may always be improved so that it is symmetric (under the equations of motion), one often disregards the spin current. Nevertheless, some recent developments seem to require information about $S^{\mu\nu\rho}$. Of particular interest to the authors of this work is the unusual nature of $\Lambda$ hyperon polarization in the quark gluon plasma \cite{STAR:2017ckg,Adam:2018ivw} which can be explained by hydrodynamic properties of the angular momentum density, see, e.g.,  \cite{Voloshin:2004ha,Becattini:2007sr,Betz:2007kg,Florkowski:2017ruc,Hongo:2021ona,Gallegos:2021bzp}. 

In order to better understand the role of the spin current in hydrodynamic behavior one may appeal to holographic duality. Holographic duality relates the dynamics of certain supersymmetric theories to the dynamics of gravity in asymptotically AdS space \cite{Maldacena:1997re,Gubser:1998bc,Witten:1998qj}. In particular, it describes the dynamics of the fluid phase of such a theory in terms of the dynamics of a black brane \cite{Bhattacharyya:2007vjd}. Its utility in describing various features of the quark gluon plasma is probably due to the robustness of hydrodynamics; coarse graining leads to a hydrodynamic theory which is agnostic of the details of particle interactions, supersymmetric or not.

Apart from the pioneering work of \cite{Gallegos:2020otk} there has been little to no work on holographic studies of the spin current. Since the spin current is sourced by torsion it is natural to assume that one would need to include dynamical torsion to the gravitational end of the holographic description. See, e.g., \cite{Gallegos:2020otk,Erdmenger:2022nhz,Erdmenger:2023hne,Aviles:2023igk}. Further, holography with torsion may provide an interesting venue for the study of condensed matter systems involving dislocations or spintronics \cite{KATANAEV19921,Furtado_1999,Mesaros:2009az,deJuan:2009ldt,Fumeron_2017,PhysRevB.93.094418,Geracie:2014mta,Randono:2010cd}. Torsion in lattice field theories was considered in \cite{Imaki:2019ite}.

An alternate method of obtaining the spin current by holographic means is by identifying the spin current operator and using the dictionary to identify its dual bulk field. The boundary value of this bulk field will source the spin current and therefore generates torsion. Our main finding is that the stress tensor multiplet of \cite{Komargodski:2010rb} (henceforth the KS multiplet) contains information about the spin current associated with the angular momentum density of supersymmetric theories---when the latter is expressed in a supersymmetric manner. We find that the spin current is given by the Hodge dual of the bottom component of the stress tensor multiplet. When the theory is conformal invariant, or, when an $R$ current exists, the spin current is given by the Hodge dual of the $R$ current. This result provides an interesting route to computing the spin current in holographic theories with $\mathcal{N}=1$ supersymmetry.

Our main result leans on an explicit computation where the stress tensor is coupled to a vierbein and an independent spin connection (or torsion). Such a coupling allows for an explicit identification of a spin current. After reviewing the notion of torsion in section \ref{S:Torsion}, we carry out the latter analysis and a variety of cross checks of it in section \ref{S:coupling}. We end this work with a discussion of various possible extensions and implications of this result \ref{S:discussion}.

\section{Torsion}
\label{S:Torsion}
We work with a vierbein $e^{a}{}_{\mu}$ and spin connection $\omega_{\mu}{}^{ab}$. We use Greek letters for coordinate indices and roman letters for tangent space indices. We use $\Gamma^{\mu}{}_{\nu\rho}$ for the connection, while $\mathring{\Gamma}^{\mu}{}_{\nu\rho}$ is the Christoffel connection
\begin{align}
	\mathring{\Gamma}^{\mu}{}_{\nu\rho} &= \frac{1}{2} g^{\mu\sigma}\left(\partial_{\nu}g_{\sigma \rho} + \partial_{\rho}g_{\sigma \nu} - \partial_{\sigma}g_{\nu\rho}\right)\, ,\\
	 \Gamma^{\mu}{}_{\nu\rho} & = \mathring{\Gamma}^{\mu}{}_{\nu\rho} + K_{\nu}{}^{\mu}{}_{\rho}\,.
\end{align}
The contorsion tensor $K_{\mu}{}^{\nu\rho}$ is related to the torsion tensor $T^{\mu}{}_{\nu\rho}$ via
\begin{equation}
	T^{\mu}{}_{\nu\rho} = K_{\nu}{}^{\mu}{}_{\rho} - K_{\rho}{}^{\mu}{}_{\nu}\,.
\end{equation}
More generally, ringed quantities will refer to those associated with a torsionless manifold in contrast to their non-ringed counterparts which encode torsion, e.g.,
\begin{align}
\begin{split}
\label{E:ringed}
	\mathring{\omega}_{\mu}{}^{ab} &= e{}^{a}{}_{\nu} \left(\partial_{\mu}e^{b \nu} + \mathring{\Gamma}^{\nu}{}_{\mu\sigma}e^{b \sigma}\right) \, ,\\
	\omega_{\mu}{}^{ab} &= \mathring{\omega}_{\mu}{}^{ab} + K_{\mu}{}^{ab}\,.
\end{split}
\end{align}
The same applies to the covariant derivative, $\mathring{\nabla}_{\mu}$ and $\nabla_{\mu}$, the Riemann tensor, $\mathring{R}^{\mu}{}_{\nu\rho\sigma}$ and $R^{\mu}{}_{\nu\rho\sigma}$ etc.

The vielbein and spin connection couple to the stress tensor $T^{\mu}{}_{a}$ and spin current $S_{\mu}{}^{ab}$ respectively, viz.
\begin{equation}
	\delta S = \int d^4x |e| \left(T^{\mu}{}_{a} \delta e^a{}_{\mu} + \frac{1}{2} S^{\mu}{}_{ab} \delta \omega_{\mu}{}^{ab} + E \cdot \delta \psi\right)\,,
\end{equation}
where $\psi$ denotes the dynamical fields and $E$ the equations of motion.
If the action is invariant under coordinate transformations $x^{\mu} \to x^{\mu} + \xi^{\mu}$ and Lorentz transformations parameterized by an infinitesimal boost parameter $\theta^{a}{}_{b}$
\begin{align}
\begin{split}
\label{E:diffboost}
	\delta e^{a}{}_{\mu} &= \xi^{\nu} \partial_{\nu} e^{a}{}_{\mu} + \partial_{\mu}\xi^{\nu} e^a{}_{\nu} - \theta^a{}_{b} e^b{}_{\mu} \, , \\
	\delta \omega_{\mu}{}^{ab}& = \xi^{\nu} \partial_{\nu} \omega_{\mu}{}^{ab} + \partial_{\mu}\xi^{\nu} \omega_{\nu}{}^{ab}  + {\nabla}_{\mu} \theta^{ab}\,,
\end{split}
\end{align}
then we find that the associated conservation equations are
\begin{align}
\begin{split}
\label{E:fullWard}
   \mathring{\nabla}_{\mu} T^{\mu\nu} &= \frac{1}{2} R^{\rho\sigma\nu\lambda}S_{\rho\lambda\sigma} - T_{\rho}{}_{\sigma}K^{\nu a b}e{}_{a}{}^{\rho}e{}_{b}{}^{\sigma}\, , \\
	\mathring{\nabla}_{\lambda}S^{\lambda}{}_{\mu\nu} &= 2 T_{[\mu\nu]} + 2 S^{\lambda\rho}{}_{[\mu}e^a{}_{\nu]} e^b{}_{\rho} K_{\lambda a b}\,.
\end{split}
\end{align}	
where  square brackets denote antisymmetrization. Later we will also use circular brackets for symmetrization,
\begin{align}
\begin{split}
	A_{[\mu\nu]} &= \frac{1}{2} \left(A_{\mu\nu} - A_{\nu\mu}\right) \,, \\
	A_{(\mu\nu)} &= \frac{1}{2} \left(A_{\mu\nu} + A_{\nu\mu}\right) \,.
\end{split}
\end{align}

We emphasize that the conservation equations in \eqref{E:fullWard} are only valid under the equations of motion. To make this point clear we rewrite them as
\begin{align}
\begin{split}
\label{E:flatWard}
	\mathring{\nabla}_{\mu}T^{\mu\nu} &= E^{\nu}\, , \\
	\mathring{\nabla}_\lambda S^{\lambda\mu\nu} &= 2 T^{[\mu\nu]} + E^{\mu\nu}\, ,
\end{split}
\end{align}
in a flat torsionless background, with $E^{\mu}$ and $E^{\mu\nu}$ proportional to the equations of motion or their derivatives. The first equality in \eqref{E:flatWard} is associated with invariance under coordinate transformations and leads to momentum conservation. The second equality is a result of local Lorentz invariance and leads to angular momentum conservation. Indeed, 
\begin{equation}
	\mathring{\nabla}_{\mu} J^{\mu\nu\rho} = 0 
\end{equation}
(under the equations of motion) follows from the second equality in \eqref{E:flatWard} if we make the identification
\begin{equation}
	J^{\mu\nu\rho} =  x^{\rho}T^{\mu\nu} - x^{\nu}T^{\mu\rho} +S^{\mu\nu\rho}\,.\label{E:Jdef}
\end{equation}

It should be pointed out that one may always add improvement terms to the stress tensor and spin current which do not affect the conservation equations in \eqref{E:flatWard}. In particular, 
defining
\begin{align}
\begin{split}
\label{E:TSprimed}
	T^{\prime\mu\nu} & = T^{\mu\nu}  + \frac{1}{2} \mathring{\nabla}_{\rho} \Psi^{\mu\nu\rho} \, , \\
	S^{\prime\mu\nu\rho} & = S^{\mu\nu\rho} + \Psi^{\mu[\nu\rho]}
\end{split}
\end{align}
with $\Psi^{\mu\nu\rho} = -\Psi^{\rho\nu\mu}$
results in 
\begin{align}
\begin{split}
\label{E:flatWardprime}
	\mathring{\nabla}_{\mu}T^{\prime\mu\nu} &= E^{\nu} \, ,\\
	\mathring{\nabla}_\lambda S^{\prime\lambda\mu\nu} &= 2 T^{\prime[\mu\nu]} + E^{\mu\nu}\,.
\end{split}
\end{align}
That is, the conservation equations obeyed by the improved stress tensor and spin current given in \eqref{E:flatWardprime} are satisfied if and only if the original conservation equation \eqref{E:flatWard} is satisfied. The improvement terms proportional to $\Psi^{\mu\nu\rho}$ do not affect the equations of motion.
Of course, one may always choose $\Psi^{\mu\nu\rho} = S^{\nu\rho\mu}+S^{\rho\nu\mu}-S^{\mu\nu\rho}$ in which case we have
\begin{align}
\begin{split}
\label{E:TSprimeprimed}
	T^{\prime\prime\mu\nu} & = T^{\mu\nu}  + \frac{1}{2}\mathring{\nabla}_\rho\left( S^{\nu\rho\mu}+S^{\rho\nu\mu}-S^{\mu\nu\rho} \right)  \, ,  \\
	S^{\prime\prime\mu\nu\rho} & = 0\,.
\end{split}
\end{align}
The resulting conservation laws are
\begin{align}
\begin{split}
\label{E:flatWardprimeprime}
	\mathring{\nabla}_{\mu}T^{\prime\prime\mu\nu} &= E^{\nu} \, , \\
	2 T^{\prime\prime[\mu\nu]}  &= E^{ \mu\nu}\,.
\end{split}
\end{align}

Note that $T^{\prime\prime \mu\nu}$ is symmetric only under the equations of motion. To obtain the improved, manifestly symmetric stress tensor we use
\begin{equation}
	T_s^{\mu\nu} = T^{\prime\prime \mu\nu} - T^{\prime\prime[\mu\nu]} \ ,
\end{equation}
which is conserved under the equations of motion and is identically symmetric
\begin{align}
\begin{split}
\label{E:flatWard2}
	\mathring{\nabla}_{\mu}T_s^{\mu\nu} &= E^{\nu} - \frac{1}{2} \mathring{\nabla}_{\mu}E^{\mu\nu}\,, \\
	2 T_s^{[\mu\nu]}  &= 0\,.
\end{split}
\end{align}
With the manifestly symmetric stress tensor, we no longer possess information regarding the symmetry of the equations of motion under local boosts in the tangent space.

In what follows we will be interested in the structure of the spin current in supersymmetric theories, with an eye towards holographic constructions \cite{Gallegos:2020otk,Erdmenger:2022nhz,Erdmenger:2023hne,Aviles:2023igk}.~\footnote{We emphasize that we are not considering supercurrents associated with higher spin fields as discussed in, e.g., ~\cite{Buchbinder:2017nuc,Koutrolikos:2017qkx,Buchbinder:2018wwg,Buchbinder:2018gle}, which may be interesting in their own right.} Our goal is to obtain a spin current compatible with supersymmetry. As we will show in the next section, such a current can be obtained by studying the structure of the stress tensor multiplet.

\section{Coupling supersymmetric theories to torsion}
\label{S:coupling}
The authors of \cite{Komargodski:2010rb} discussed an $\mathcal{N}=1$ four dimensional stress tensor multiplet $\mathbf{S}_{\mu}$ satisfying the conservation law
\begin{subequations}
\label{E:SKdef}
\begin{equation}
	\overline{D}^{\dot{\alpha}} \sigma_{\alpha\dot{\alpha}}^a \delta_{a}{}^{\mu} \mathbf{S}_{\mu} = -\frac{1}{2} D_{\alpha}\mathbf{X} -\frac{1}{2} {\boldsymbol{\chi}}_{\alpha} \, ,
\end{equation}
with
\begin{align}
\begin{split}
\label{E:Xchi}
	\overline{D}_{\dot{\alpha}} \mathbf{X} & = 0 \, , \\
	\overline{D}_{\dot{\alpha}}{\boldsymbol{\chi}}_{\alpha} &= \overline{D}_{\dot\alpha}\overline{{\boldsymbol{\chi}}}^{\dot{\alpha}} - D^{\alpha} {\boldsymbol{\chi}}_{\alpha} = 0 \,.
\end{split}
\end{align}
\end{subequations}
Our conventions for supersymmetry can be found in appendix \ref{A:SUSY}.~\footnote{
Although potentially interesting to pursue, we will not consider the alternative 4D $\mathcal{N}=1$ theory of complex linear superfields~\cite{Gates:1980az,Koci:2016rqf}
}
We will refer to $\mathbf{S}_{\mu}$ as the Komargosdki Seiberg  (KS) multiplet. Note that we have introduced a (flat space) vierbein $e_a{}^{\mu} = \delta_a{}^{\mu}$ in \eqref{E:SKdef} to emphasize the index structure of the stress tensor multiplet. In this work we are only interested in coupling the stress tensor to linear fluctuations of the vierbein in which case expression \eqref{E:SKdef} remains unchanged.

As discussed in \cite{Komargodski:2010rb} the components of $\mathbf{S}_{\mu}$ can be derived from the conservation law given in \eqref{E:SKdef}. The bosonic terms of $\mathbf{S}_{\mu}$ are given by
\begin{align}
\begin{split}
\label{E:SKcomponents}
    S_{1\,\mu} & = J_{\mu} \, ,\\
    -\frac{1}{4} S_{\theta^2\mu} & = \frac{i}{2} \partial_{\mu}\bar{X}_1 \, , \\
    -\frac{1}{4}\bar{S}_{\bar{\theta^2}\mu} & = -\frac{i}{2} \partial_{\mu}X_1 \, ,\\
    \frac{1}{2} S_{\theta\bar{\theta}\mu a} & =  \left( 2 T_{s\,\mu \nu} - \eta_{\mu\nu} Z - \frac{1}{2} \varepsilon_{\mu\nu\rho\sigma}\left(\partial^{\rho} J^{\sigma} + F^{\rho\sigma}\right) \right) \delta{}_{a}{}^{\nu} \, ,\\
    \frac{1}{16} S_{\theta^2 \bar{\theta}^2} &= \frac{1}{2} \partial_{\mu}\partial^{\nu} J_{\nu} - \frac{1}{4} \partial^2 J_{\mu}\, ,
\end{split}
\end{align}
together with
\begin{align}
\begin{split}\label{E:X and Chi def}
    X_{\theta^2} &= -4\left(Z + i \partial^{\nu}J_{\nu}\right) \, , \\
    \chi_{\theta \alpha}{}^{\beta} &=-\delta_{\alpha}^{\beta} \left(-4 T_s^{\mu}{}_{\mu}+6 Z\right) + 2 i \left(\sigma^{\rho}\bar{\sigma}^{\sigma}\right)_{\alpha}{}^{\beta} F_{\rho\sigma}\,.
\end{split}
\end{align}
where $Z$ is real, $T_s^{\mu\nu}$ is conserved and symmetric, and $F_{\mu\nu}$ are the components of a closed two-form.  The manifestly symmetric tensor $T_s^{\mu\nu}$ is identified with the symmetric stress tensor.

We would like to identify possible spin currents consistent with supersymmetry. To this end we consider the non trivial commutators involving supersymmetry, momentum and angular momentum generators
\begin{align}
\begin{split}
	-i \left[P_{\lambda},\,M^{\nu\rho}\right]&= \delta^{\rho}_{\lambda} P^{\nu} - \delta^{\nu}_{\lambda} P^{\rho}\,, \\
	i\left[M^{\lambda\eta}\,,M^{\nu\rho}\right]&=\eta^{\nu\lambda}M^{\rho\eta}-\eta^{\rho\lambda}M^{\nu\eta}-\eta^{\nu\eta}M^{\rho\lambda}+\eta^{\rho\eta}M^{\nu\lambda}\,, \\
	-i \left[Q,\,M^{\nu\rho}\right] & =  \sigma^{\nu\rho} Q\,,
\end{split}
\end{align}
where $\sigma^{\mu\nu}$ is defined in \eqref{E:sigmaDef}. To obtain the associated commutators for the conserved currents $T^{\mu\nu}$ and $J^{\mu\nu\rho}$ we use $P^{\nu} = \int T^{\mu\nu}d\Sigma_{\mu}$ and $M^{\nu\rho} = \int J^{\mu\nu\rho} d\Sigma_{\mu}$ with $d\Sigma_{\mu}$ a spatial volume element.
(We remind the reader that the generators of momentum $P^{\nu}$ and of angular momentum $M^{\nu\rho}$ are associated with the existence of Killing vectors related to translation invariance and rotation invariance, which can be written in the form $\xi^{\mu}_{(\alpha)}  = \delta^{\mu}_{\alpha}$ and $\xi^\mu_{\omega}=\omega^\mu{}_\rho x^\rho$. )
The conservation laws, $\mathring{\nabla}_{\mu}T^{\mu\nu}=0$ and $\mathring{\nabla}_{\mu}J^{\mu\nu\rho}=0$, imply that $P^{\mu}$ and $M^{\nu\rho}$ are, in a sense, independent of the spatial section (or, that they are topological). 
Thus,
\begin{align}
\begin{split}
\label{E:algebracurrents}
	-i \left[P^{\lambda},\,J^{\mu\nu\rho}\right]&= \eta^{\rho \lambda} T^{\mu\nu} - \eta^{\nu \lambda} T^{\mu\rho} + \mathcal{S}_{PJ}^{\mu\lambda \nu\rho} \,, \\
	i \left[M^{\nu\rho},\,T^{\mu\lambda}\right]&= \eta^{\rho \lambda} T^{\mu\nu} - \eta^{\nu \lambda} T^{\mu\rho} + \mathcal{S}_{MT}^{\mu\lambda \nu\rho} \,, \\
	i\left[M^{\lambda\eta}\,,J^{\mu\nu\rho}\right]&=\eta^{\nu\lambda}J^{\mu\rho\eta}-\eta^{\rho\lambda}J^{\mu\nu\eta}-\eta^{\nu\eta}J^{\mu\rho\lambda}+\eta^{\rho\eta}J^{\mu\nu\lambda} + \mathcal{S}_{MJ}^{\mu\lambda\eta\nu\rho}\,, \\
	-i \left[Q,\,J^{\mu\nu\rho}\right] & =  \sigma^{\nu\rho} J_Q^{\mu} + \mathcal{S}_{QJ}^{\mu\nu\rho} \,.\\
\end{split}
\end{align}
where $J_Q^{\mu}$ is the super current and the various $\mathcal{S}$ are referred to as Schwinger terms: they are conserved and their integral over a spatial section (orthogonal to the $\mu$ index) vanishes. 
In addition, we have dropped expressions which vanish under the equations of motion.

The angular momentum density, $J^{\mu\nu\rho}$ is composed of an orbital component  and a spin component, c.f., \eqref{E:mainJ}. We refer to a pair $T^{\mu\nu}$, $S^{\mu\nu\rho}$ as a valid pair if they combine to form an angular momentum density satisfying the algebra given by \eqref{E:algebracurrents}. 
Let us consider, for instance, $T^{\mu\nu} = T_s^{\mu\nu}$ and $S^{\mu\nu\rho}=0$ where $T_s^{\mu\nu}$ is defined in the bosonic components of the supercurrent in equation \eqref{E:SKcomponents} and is assumed to be conserved due to translation invariance. Since $P^{\mu}$ and $M^{\nu\rho}$ generates translations and rotations respectively, we have
\begin{equation}\label{E:PandMaction}
	i a^{\mu} [P_{\mu},\, O] =  \pounds_{a^{\mu}\xi_{(\mu)}}O \,,
	\qquad
	\frac{1}{2} i \omega_{\lambda\eta} [M^{\lambda\eta},\,O] = \pounds_{\xi_{\omega}} O \, ,
\end{equation}
for an $O$ which does not depend explicitly on the coordinates and constant $a$ and $\omega$.
Thus, $T^{\mu\nu} = T_s^{\mu\nu}$ and $S^{\mu\nu\rho}=0$ are compatible with the first three equalities of the algebra given by \eqref{E:algebracurrents}. A somewhat tedious computation implies that 
\begin{equation}
	i [Q\,,T_s^{\mu\nu}] =
	\partial_{\lambda}\left(\frac{1}{2} \sigma^{\mu\lambda}J_{Q}^{\nu}
	+\frac{1}{2}\sigma^{\nu\lambda}J_{Q}^{\mu}
	-\frac{1}{2}\sigma^{\nu\mu}J_{Q}^{\lambda}\right)\,.
\end{equation}
Thus, the last equality of the algebra given by \eqref{E:algebracurrents} is satisfied and $T^{\mu\nu} = T_s^{\mu\nu}$ and $S^{\mu\nu\rho}=0$ are a valid pair as expected. 

Next suppose that $T^{\mu\nu} = T_s^{\mu\nu} + \frac{1}{2} \mathring{\nabla}_{\rho} \Psi_1^{\mu\nu\rho}$ and $S^{\mu\nu\rho} = \Psi_1^{\mu[\nu\rho]}$ where $\Psi_1^{\mu\nu\rho}$ is the bottom component of a multiplet $\boldsymbol{\Psi}^{\mu\nu\rho}$ which does not depend explicitly on the spatial coordinates. A straightforward computation implies that the first three equalities of the algebra given by \eqref{E:algebracurrents} will still be satisfied. Further, in flat spacetime, supersymmetry dictates that 
\begin{equation}
	[Q,\,T^{\mu\nu}] = [Q,\,T^{\mu\nu}_s] + \mathring{\nabla}_{\lambda} \Psi^{\mu\nu\lambda}_{\theta} \, ,
\end{equation}
which leads to the fourth equality of the algebra given by \eqref{E:algebracurrents}.

We have found that, if our considerations involve only the supersymmetry algebra, any spin current can be added to the stress tensor as an improvement term; we may always declare it to be the bottom component of a spin current multiplet whose other components can be generated by acting on it consecutively with $Q$ and $\bar{Q}$. The background source which couples to the spin current is the spin connection. Thus, for consistency, the spin connection should sit in the top component of a multiplet.  To proceed further, we need to study the vierbein multiplet.

The authors of \cite{Komargodski:2010rb} provided a detailed discussion of the coupling between the stress tensor multiplet and the graviton multiplet $ \mathbf{H}^{\mu}$. As discussed in the previous section, in coupling the stress tensor to the metric it is difficult to identify its symmetric and antisymmetric components and, accordingly, the spin current. In what follows we will follow \cite{Komargodski:2010rb}  but couple the stress tensor multiplet to a vierbein multiplet which will allow us to identify the antisymmetric components of the stress tensor. As it turns out, the spin connection also appears in the vierbein multiplet. As a result we can check that our expression for the antisymmetric component of the stress tensor is compatible with an explicit computation of the spin current. As an additional check of our construction we demonstrate that the resulting linearized equations for the vierbein and spin connection are commensurate with the linearized equations of supergravity.

Let 
\begin{equation}
	e{}^{a}{}_{\mu} = \delta{}^{a}{}_{\mu} +  h{}^{a}{}_{\mu}\,.
\end{equation}
To linear order in $h^{a}{}_{\mu}$ the coupling of the stress tensor to vierbein takes the form $T^{\mu}{}_{a} h{}^{a}{}_{\mu}$. Since the stress tensor appears in the $\theta \bar{\theta}$ component of $\mathbf{S}_{\mu}$ we expect that $h{}^{a}{}_{\mu}$ appear in the same component of a real vierbein multiplet $\mathbf{E}_{\mu}$,
\begin{equation}
	\delta^{b}{}_{\mu}E_{\theta\bar{\theta}}{}^{\mu}{}_{a} = \delta_{a}{}^{\mu}h{}^{b}{}_{\mu} + \hbox{other (possible) contributions}\,.
\end{equation}
The transformation rules for the vierbein are given in \eqref{E:diffboost} and we expect that the entire vierbein multiplet will transform such that,
\begin{equation}
	\mathbf{E}_{\mu} \to  \mathbf{E}_{\mu} + \delta_L\mathbf{E}_{\mu}\,.
\end{equation}
where $\delta_L \mathbf{E}_{\mu}$ encodes \eqref{E:diffboost}; the subscript $L$ is to remind us that we are not carrying a general variation but a variation associated with a coordinate transformation. In order to generate the $\pounds_{\xi}e{}^a{}_{\mu}$ term appearing in \eqref{E:diffboost} in the $\theta \bar{\theta}$ component of $\delta_L \mathbf{E}_{\mu}$ we need that, at the very least, 
\begin{equation}
\label{E:coordinateT}
	\sigma^{a}_{\alpha\dot{\alpha}} \delta_{a}{}^{\mu} \delta_L \mathbf{E}_{\mu} =  D_{\alpha} \bar{\mathbf{L}}_{\dot{\alpha}} - \bar{D}_{\dot\alpha} {\mathbf{L}}_{{\alpha}} \,.
\end{equation}
Indeed, if we denote 
\begin{align}
\begin{split}
\label{E:Ldef}
	\bar{L}_{\theta}^{a} &= \frac{1}{2} \zeta^{a} + \frac{1}{4} i \xi^{a} \, ,\\
	{\bar{L}_{\bar{\theta}}}{}_{\dot{\alpha}}{}^{\dot{\beta}}&=4i{\bar{\theB}}_{\dot{\alpha}}{}^{\dot{\beta}}\, , \\	
	\bar{L}_{\theta^{2}\bar{\theta} \dot{\alpha}}{}^{\dot{\beta}} & =4i {\bar{\theA}}_{\dot{\alpha}}{}^{\dot{\beta}}-i\delta_a{}^{\mu}\bar{\sigma}^{a\,\dot{\beta}\alpha}\sigma_{\alpha\dot{\alpha}}^{b}\partial_{\mu}\left(2\zeta_{b}+i\xi_{b}\right)\, , \\
	\bar{L}_{\bar{\theta}^{2}\theta}^{a}&=\bar{\theC}^{a}+2\bar{\sigma}^{a\dot{\alpha}\alpha}\sigma_{\alpha\dot{\beta}}^{b}{\delta}_b{}^{\nu}\partial_{\nu}{\bar{\theB}}_{\dot{\alpha}}{}^{\dot{\beta}}\, ,
\end{split} 
\end{align}
then the bosonic components of $\delta_L \mathbf{E}_{\mu}$ transform as
\begin{align}
\begin{split}
\label{deltaE}
	\delta_L E_{1\mu} &= \delta^a{}_{\mu} \zeta_{a} \, ,\\
	\delta_L E_{\theta \bar{\theta}}{}_{\mu b} &= \delta^a{}_{\mu}\theA_{a b} + \delta_{b}{}^{\nu}\delta^{a}{}_{\mu}\partial_{\nu}\xi_{a} + \delta^a{}_{\mu} \eta_{ab} \hbox{Im} \bar{\theA}_{\dot{\alpha}}{}^{\dot{\alpha}}\, , \\
	\delta_L E_{\theta^2 \bar{\theta}^2\mu} &= 4 \partial_{\mu} \hbox{Re} \bar{\theA}_{\dot{\alpha}}{}^{\dot{\alpha}} + 4 \delta^a{}_{\mu} \partial^2 \zeta_{a} + 2 \delta^a{}_{\mu} \varepsilon_a{}^{bcd}\delta_{b}{}^{\nu}\partial_{\nu}\theA_{dc}\, ,\\
	\delta_L E_{\bar{\theta}^{2}\mu} &= \bar{\theC}_{a}\delta^a{}_{\mu} \,.
\end{split}
\end{align}

At this point it is useful to contrast a generic KS multiplet with various edge cases. In the special case where ${\boldsymbol{\chi}}_{\alpha}=0$ the stress tensor multiplet is referred to as the Ferrara Zumino (FZ) multiplet and when $\mathbf{X}=0$ it is referred to as the R multiplet since its bottom component denotes the $R$ current. In a conformal theory it is possible to set both ${\boldsymbol{\chi}}_{\alpha}$ and $\mathbf{X}$ to zero. In order to avoid confusion we will, from now on, use $\mathbf{S}_{\mu}$ for the stress tensor multiplet, only when ${\boldsymbol{\chi}}_{\alpha} \neq 0$ and $\mathbf{X} \neq 0$. Otherwise we will denote the stress tensor multiplet by $\mathbf{T}_{\mu}$ if it is the FZ multiplet and by $\mathbf{R}_{\mu}$ if it is the $R$ multiplet. Both $\mathbf{R}_{\mu}$ and $\mathbf{T}_{\mu}$ can be used to describe the stress tensor multiplet in a conformal theory, in which case we use $\mathbf{C}_{\mu}$.

\subsection{The Ferrara Zumino multiplet}
The components of the FZ multiplet can be read off from the components of the KS multiplet given in  \eqref{E:SKcomponents} after setting $\boldsymbol{\chi}_{\alpha}=0$, viz., $F_{\mu\nu}=0$ and $6 Z =  4 T^{\mu}{}_{\mu}$. In the FZ multiplet, $\mathbf{L}_{\alpha}$ is further constrained by the linear coupling of the vierbein to the stress tensor,
\begin{equation}
\label{E:FZcoupling}
	\mathcal{L}_{int} = -\frac{1}{8} \int d^2\theta d^2\bar{\theta} \mathbf{T}^{\mu} \mathbf{E}_{\mu}\,,
\end{equation}
In order that the interaction in \eqref{E:FZcoupling} be invariant under the transformation \eqref{E:coordinateT}, it must be the case that
\begin{equation}
\label{E:FZconstraint}
	\bar{D}^2 D^{\alpha}\mathbf{L}_{\alpha} = 0\,.
\end{equation}
The constraint \eqref{E:FZconstraint} implies that the bosonic components of $L_{\alpha}$ are constrained such that
\begin{align}
\begin{split}
\label{E:FZconstraintcomponents}
	\bar{\theA}_{\dot{\alpha}}{}^{\dot{\alpha}} &= -2\partial_{\lambda}\zeta^{\lambda} -i \partial_{\lambda}\xi^{\lambda} \, ,\\
	\delta_a{}^{\mu}\partial_{\mu}  \bar{\theC}^a & = 0 \,.
\end{split}
\end{align}
Thus
\begin{align}
\begin{split}
\label{E:FZvariation}
	\delta{}_{a}{}^{\mu} \delta_L E_{1\mu} & = \zeta_{a}\, ,\\
	\delta{}_{a}{}^{\mu}  \delta_L E_{\theta\bar{\theta}\mu b} & =\theA_{ab}+\left(\delta_{a}^{c}\delta_{b}^{d}-\eta^{dc}\eta_{a b}\right)\delta_d{}^{\mu}\partial_{\mu}\xi_{c}\,, \\
	 \delta{}_{a}{}^{\mu} \delta_L E_{\theta^{2}\bar{\theta}^{2}\mu} & = -8 \delta{}_{a}{}^{\mu} \delta_b{}^{\nu} \partial_{\mu} \partial_{\nu} \zeta^b + 4  \partial^2 \zeta_{a} + 2  \varepsilon_a{}^{bcd}\delta_{b}{}^{\nu}\partial_{\nu}\theA_{dc}\,.
\end{split}
\end{align}
We can now interpret the various bosonic components of $\mathbf{E}_{\mu}$ according to their transformation properties. Comparing the transformation of the components in \eqref{E:FZvariation} to the transformation of the vierbein and spin connection in \eqref{E:diffboost} we find
\begin{align}
\begin{split}
\label{E:FZcomponents}
	\delta{}^{a}{}_{\mu}E_{\theta\bar{\theta}}{}^{\mu}{}_{b} &= \delta{}_{b}{}^{\mu}h{}^{a}{}_{\mu} - \delta^{a}{}_{b} h{}^{c}{}_{\nu}\delta_c{}^{\nu}\,, \\
	E_{\theta^2 \bar{\theta}^2\mu} &= 2 \varepsilon_\mu{}^{\lambda\nu\rho}\delta^a{}_\nu\delta^b{}_\rho\omega_{\lambda ab} - 8 \partial_{\mu}\partial_{\lambda}E_1^{\lambda} + 4 \partial^2 E_{1\mu}\,,
\end{split}
\end{align}
where we have interpreted $\xi^{\mu}=\delta_a{}^{\mu}\xi^{a}$ as an infinitesimal diffeomorphism and $\theA^{ab}$ as an infinitesimal Lorentz transformation (denoted by $\theta^{ab}$ in the previous section).
We can use $\zeta$ to gauge away the unphysical field $E_1^{\mu}$. Likewise, the transformation rules \eqref{E:FZvariation} imply that $E_{\theta^2}^{\mu}$ and $\bar{E}_{\bar{\theta}^2}^{\mu}$ are not gauge invariant, but that
\begin{equation}
\label{E:FZtheta2}
	\partial_{\mu} E_{\theta^2}^{\mu} = \Lambda\, ,
	\qquad
	\partial_{\mu} E_{\bar{\theta^2}}^{\mu} = \bar{\Lambda}\, ,
\end{equation}
can not be gauged away.
As should be clear from \eqref{E:FZcomponents} the Hodge dual of the independent spin connection appears in the top component of the gravity multiplet. 

Integrating the interaction in \eqref{E:FZcoupling} over superspace and using the vierbein superfield components in \eqref{E:FZcomponents} and \eqref{E:FZtheta2} we find that the bosonic components of $\mathcal{L}_{int}$ read
\begin{equation}
\label{E:LintFZ}
	\mathcal{L}_{int} = -\frac{1}{4} \varepsilon_{\mu }{}^{\sigma \nu\rho}\delta{}^{a}{}_{\nu} \delta{}^{b}{}_{\rho} \omega_{\sigma ab}J^{\mu} + h{}^{a}{}_{\mu}\delta_{a}{}^{\rho}\eta_{\rho\nu} \left(T_s^{\mu\nu} + \frac{1}{4}\varepsilon^{\mu\nu\tau\sigma}\partial_{\tau}J_{\sigma}\right) - \frac{i}{4} \left(\bar{X}_1 \bar{\Lambda} -  X_1 {\Lambda} \right)\,.
\end{equation}
Varying the interaction Lagrangian \eqref{E:LintFZ} with respect to the spin connection gives us the spin current 
\begin{align}
    \label{E:SFZ}
    S_{\mu}{}^{ab} = \frac{1}{2} \varepsilon_{\mu}{}^{\sigma \nu\rho}\delta{}^{a}{}_{\nu} \delta{}^{b}{}_{\rho} J_{\sigma} \, .
\end{align}
This is the main result we were after: the spin current is the Hodge dual of the bottom component of the FZ multiplet.
In the case of a conformal theory $J^{\mu}$ can be identified with the $R$ current. We will discuss this further in the next section. As a mild check on \eqref{E:SFZ} we can evaluate the stress tensor
by varying the interaction Lagrangian \eqref{E:LintFZ} with respect to the vielbein\,,
\begin{equation}
\label{E:TFZ}
    T^{\mu}{}_{a} = \delta_{a}{}^{\rho}\eta_{\rho\nu} \left(T_s^{\mu\nu} + \frac{1}{4}\varepsilon^{\mu\nu\tau\sigma}\partial_{\tau}J_{\sigma}\right)\,,
\end{equation}
and verify that the conservation equations in \eqref{E:flatWard} are satisfied.

As an additional consistency check of our construction, we can evaluate the linearized Einstein equations associated with $\mathbf{E}_{\mu}$. A standard method of obtaining the equations of motion for the vierbein is to construct a multiplet $\mathbf{N}_{FZ}^{\mu}$, which is linear in $\mathbf{E}_{\mu}$, contains at most two derivatives, and is inert under the superfield transformation in \eqref{E:coordinateT}. The kinetic term for the vierbein multiplet is then given by the Lagrangian
\begin{equation}
\label{E:kineticFZ}
	\mathcal{L}_{kin}  =\int d^{4}\theta \mathbf{E}_{\lambda}\mathbf{N}_{FZ}^{\lambda}\,.
\end{equation}
The unique $\mathbf{N}_{FZ}^{\mu}$ which can be constructed with the above requirements is
\begin{equation}
\label{E:NFZ}
	\mathbf{N}_{FZ}^{\lambda}  =\frac{1}{24}\varepsilon^{\lambda\sigma\nu\rho}\delta^b{}_{\nu} \bar{\sigma}_{b}^{\dot{\delta}\delta}\left[\bar{D}_{\dot{\delta}},D_{\delta}\right]\partial_{\sigma}\mathbf{E}_{\rho}-\frac{1}{24}\left(\frac{1}{2}\left\{ \bar{D}^{2},D^{2}\right\} -4\partial^{2}\right)\mathbf{E}^{\lambda}-\frac{1}{3}\partial^{\lambda}\partial_{\rho}\mathbf{E}^{\rho}+\frac{1}{6}\partial^{2}\mathbf{E}^{\lambda}\,.
\end{equation}
An explicit computation yields the bosonic Lagrangian
\begin{equation}
	\mathcal{L}_{kin} = \left(\omega_{[\mu\nu\rho]}-{\mathring\omega}_{[\mu\nu\rho]}\right)^2 -h{}^{a}{}_{\mu} \delta_a{}^{\nu}\eta_{\nu\rho} \hbox{Ein}\left[h\right]^{\mu\rho}+\frac{2}{3}\Lambda \bar{\Lambda} \,,\label{E:Lkin}
\end{equation}
 where
\begin{align}
\begin{split}
\label{E:linEin}
	\hbox{Ein}\left[h\right]^{\lambda\mu}=\partial^{2}h^{(\mu\lambda)}-\partial_{\nu}\partial^{\mu}h^{(\nu\lambda)}-\partial_{\nu}\partial^{\lambda}h^{(\mu\nu)}+\eta^{\mu\lambda}\partial_{\nu}\partial_{\rho}h^{\nu\rho}-\eta^{\lambda\mu}\partial^{2} h +\partial^{\lambda}\partial^{\mu} h \, ,
\end{split}
\end{align}
and we have defined 
\begin{equation}
\label{E:defhmunu}
	h^{\mu\nu} = \eta^{\mu\rho}\delta_a{}^{\nu} h{}^a{}_{\rho} \,\, ,
	\qquad
	h = h{}^{a}{}_{\mu} \delta_a{}^{\mu}\,\, ,
    \qquad
    \omega_{\mu\nu\rho}=\delta^a{}_\nu\delta^b{}_\rho\omega_{\mu ab}\,\,,
    \qquad
    \mathring{\omega}_{\mu\nu\rho}=\delta^a{}_\nu\delta^b{}_\rho\mathring{\omega}_{\mu ab}\,.
\end{equation}
As expected, 
$\hbox{Ein}[h]^{\mu\nu}=0$ is precisely the linearized Einstein equations and 
the equation of motion for the spin connection is 
\begin{equation}\label{eq:linear_spin_connection_eom}
	\omega^{[\mu\nu\rho]} = \mathring{\omega}^{[\mu\nu\rho]}\,.
\end{equation}
In fact, the bosonic Lagrangian \eqref{E:Lkin} can be identified with an expansion of $-\left|e\right|\left(R\left(\omega\right)-\frac{2}{3} |\Lambda|^2\right)$ 
 to quadratic order in the fields (and assuming that the contorsion is totally antisymmetric).

\subsection{The Komargodski Seiberg  multiplet}
The Komargodski Seiberg  multiplet $\mathbf{S}_{\mu}$ is characterized by \eqref{E:SKdef} with $\mathbf{X} \neq 0$ and ${\boldsymbol{\chi}}_{\alpha} \neq 0$. The coupling of $\mathbf{S}^{\mu}$ to the vierbein multiplet $\mathbf{E}_{\mu}$ takes the form
\begin{equation}
\label{E:SKcoupling}
    \mathcal{L}_{int} = -\frac{1}{8} \int d^2\theta d^2 \bar{\theta} \mathbf{S}^{\mu} \mathbf{E}_{\mu}\,,
\end{equation}
similar to the coupling of the FZ multiplet to vierbein multiplet in \eqref{E:FZcoupling}. If the superfield integral \eqref{E:SKcoupling} is to be invariant under the superfield transformations in \eqref{E:coordinateT} we need that
\begin{subequations}
\label{E:SKconstraints}
\begin{equation}
\label{E:SKconstraintFZ}
    \bar{D}^2 D^{\alpha} L_{\alpha} = 0\, ,
\end{equation}
and
\begin{equation}
    \label{E:SKconstraintR}
    \bar{D}_{\dot{\alpha}} D^2 \bar{L}^{\dot{\alpha} }= D_{\alpha} \bar{D}^2 L^{\alpha} \,.
\end{equation}
\end{subequations}
The constraint \eqref{E:SKconstraintFZ} is identical to the constraint \eqref{E:FZconstraint} and leads to the same constraint on the components \eqref{E:FZconstraintcomponents}. The additional constraint \eqref{E:SKconstraintR} implies that the bosonic components of $\mathbf{L}_{\alpha}$ satisfy 
\begin{align}
\begin{split}
\label{E:Rconstraintcomponents}
    \text{Im}\bar{\theA}_{\dot{\alpha}}{}^{\dot{\alpha}} &=0\,,   \\
    \delta_a{}^{\mu}\delta_b^{\nu} \theA^{ab} &=2\partial^{[\mu} w^{\nu]} \,.
\end{split}
\end{align}

The constraints \eqref{E:FZconstraintcomponents} together with \eqref{E:Rconstraintcomponents} read 
\begin{equation}
\label{E:badconstraint}
    \delta_a{}^{\mu} \partial_\mu\xi^a = 0\,.
\end{equation}
If $\xi^{\mu} = \delta_a{}^{\mu}\xi^a$ is to be interpreted as an infinitesimal coordinate transformation, \eqref{E:badconstraint} implies that all coordinate transformations must be divergenceless.
Likewise, if we interpret $\theA^{ab}$ as an infinitesimal boost parameter then \eqref{E:Rconstraintcomponents} implies that local boosts are exact.

As discussed in \cite{Komargodski:2010rb} one may resolve this problem by adding compensator fields which relax either one of \eqref{E:SKconstraints}. In particular, one may add a chiral compensator field ${\boldsymbol{\lambda}}_{\alpha}$ such that
\begin{equation}
\label{E:SKintwithlambda}
	\mathcal{L}_{int} = -\frac{1}{8} \int d^2\theta d^2 \bar{\theta} \mathbf{S}_{\mu} \mathbf{E}^{\mu} - \frac{1}{48}  \left(\int d^2\theta {\boldsymbol{\lambda}}^{\alpha}{\boldsymbol{\chi}}_{\alpha} + c.c.\right)\,.
\end{equation}
In this case the interaction Lagrangian \eqref{E:SKintwithlambda} can be made invariant under the transformation in \eqref{E:coordinateT} such that only \eqref{E:SKconstraintFZ} is satisfied and  \eqref{E:SKconstraintR} is replaced by the transformation rule
\begin{equation}
\label{E:SKlambdaconstraint}
	\delta_L {\boldsymbol{\lambda}}^{\alpha} = \frac{3}{4} \bar{D}^2 \mathbf{L}^{\alpha}\,.
\end{equation}
From the conservation law and the conditions on the superfield $\boldsymbol{\chi}$ in \eqref{E:Xchi}, the coupling \eqref{E:SKintwithlambda} is also invariant under
\begin{equation}
\label{E:SKgauge}
      \delta_{V} {\boldsymbol{\lambda}}^{\alpha} = \frac{3}{4}i \bar{D}^2 D^{\alpha} {\boldsymbol{\theL}}\,,
\end{equation}
with ${\boldsymbol{\theL}}$ a real superfield.

With ${\boldsymbol{\lambda}}_{\alpha}$ as a compensator field, the constraint \eqref{E:SKconstraintFZ} on $\mathbf{L}_{\alpha}$ is identical to \eqref{E:FZconstraint}. Thus, the components of the vierbein multiplet $\mathbf{E}_{\mu}$ are identical to those found with the FZ multiplet in \eqref{E:FZcomponents}. The additional bosonic degrees of freedom associated with the $\theta$ component of ${\boldsymbol{\lambda}}_{\alpha}$ can be parameterized by 
\begin{equation}
	\lambda_{\theta}{}^\alpha{}_\beta = \frac{1}{2} \delta_{\beta}^{\alpha} \lambda_{\theta} +\frac{1}{2} 
     i \sigma_{ab}{}^{\alpha}{}_{\beta} \lambda_{\theta}^{ab}\,.
\end{equation}

In this notation the transformation rule \eqref{E:SKlambdaconstraint} reads
\begin{align}
\begin{split}
\label{E:deltaLlambdaSK}
	\delta_L\lambda_{\theta}&=3i\delta{}_{a}{}^{\mu}\left(2\partial_{\mu}\zeta^{a}+i\partial_{\mu}\xi^{a}\right)\,,\\
        \delta_L\lambda_{\theta}^{ab}&=\frac{3}{2}\varepsilon^{abcd}\theA_{cd}\,,
\end{split}
\end{align}
and the transformation rule \eqref{E:SKgauge} reads
\begin{align}
    \delta_{\theL} \lambda_{\theta}  &= iv \, ,\\
    \delta_{\theL} \lambda_{\theta}^{ab} & = \frac{3}{2}\partial^{\mu}\delta^{[a}{}_{\mu}v^{b]}\,,
\end{align}
where 
\begin{equation}
	    v=-3\left(\frac{1}{16}V_{\theta^2\bar\theta^2}-\frac{1}{2}\partial^2V_1\right),\qquad
	    v^{\mu} = \delta_a{}^{\mu} V_{\theta\bar{\theta}}^a\,.
\end{equation}

The transformation rule \eqref{E:deltaLlambdaSK} for $\delta_L \lambda_{\theta}^{ab}$ is similar to the transformation rule for the Hodge dual of the vierbein
\begin{equation}
    \delta_L \left(\delta{}_{c}{}^{\mu}\varepsilon^{abc}{}_{d}h^d{}_{\mu}\right) 
    =
    -\varepsilon^{abcd}\left(\theA_{cd}-\delta{}_{c}{}^{\sigma}\partial_{\sigma}\xi_{d}\right)\,.
\end{equation}
In order to obtain \eqref{E:deltaLlambdaSK} we conclude that
\begin{equation}
\label{E:gotlambdathetaab}
    \lambda_{\theta}^{ab}=\frac{3}{2}\delta{}_{c}{}^{\mu}\varepsilon^{abc}{}_{d}\left(\delta{}^{d}{}_{\nu}B_{\mu}{}^{\nu}-h^d{}_{\mu}\right)\, ,
\end{equation}
where $B_{\mu\nu}$ is (an antisymmetric) Kalb-Rammond field satisfying
\begin{align}
\begin{split}
\label{E:Brules}
    \delta_L B^{\mu\nu}&=\delta{}_{c}{}^{[\nu}\partial^{\mu]}\xi^{c}\,, \\  \delta_{\theL}B_{\mu\nu} &= \partial_{\mu}v_{\nu} - \partial_{\nu}v_{\mu}\,,
\end{split}
\end{align}
The transformation law $B_{\mu\nu} \to B_{\mu\nu} + \delta_{L} B_{\mu\nu}$ does not coincide with a coordinate transformation of $B^{\mu\nu}$. The latter cannot be observed within the perturbative formalism we are working in since $B_{\mu\nu}$ represents a perturbation of the Kalb-Ramond field around $B_{\mu\nu}=0$. Note, however, that $B_{\mu\nu} \to B_{\mu\nu} + \delta_{\theL} B_{\mu\nu}$ is commensurate with a gauge transformation with parameter $\frac{1}{2} \delta^c{}_{\mu}\xi_c$. 
In addition to providing the correct transformation rules \eqref{E:deltaLlambdaSK} for $\lambda_{\theta}^{ab}$, $B_{\mu\nu}$ also ensures that ${\boldsymbol{\lambda}}_{\alpha}$ is a chiral superfield. 

The transformation rule \eqref{E:deltaLlambdaSK} for $\lambda_{\theta}$ is compatible with 
\begin{equation}
    \delta_L  \left(2 \partial_{\mu}E_1^{\mu} + i h \right) = 
    \delta_{a}{}^{\mu}\left(2\partial_{\mu}\zeta^{a}+i\partial_{\mu}\xi^{a}\right)\,, 
\end{equation}
(with $h=\delta_a{}^{\mu}h^{a}{}_{\mu}$)
suggesting
\begin{align}
\begin{split}
\label{E:gotlambdatheta}
    \hbox{Im} \lambda_{\theta} &= \Xi + 6 \partial_{\mu} E_1^{\mu} \,,\\
    \hbox{Re} \lambda_{\theta} &= \frac{1}{2} \left(\Phi - 6 h\right)\, ,\\
\end{split}
\end{align}
where $\Xi$ encodes the gauge transformation for $\lambda_{\theta}$,
\begin{align}
\begin{split}
    \delta_{\theL}\Xi &=v \, ,\\
    \delta_{\theL} \Phi &= 0 \, ,
\end{split}
\end{align}
and both $\Xi$ and $\Phi$ are needed in order to ensure that ${\boldsymbol{\lambda}}_{\alpha}$ is a chiral superfield.\footnote{The transformation law for $h$ is completely determined by the structure of the $\mathbf{E}_{\mu}$ multiplet in which it sits. In the absence of $\Phi$, the transformation rule for $\hbox{Re}\lambda_{\theta}$ would be completely determined by the transformation rule for $h$ which would lead to a non chiral ${\boldsymbol{\lambda}}_{\alpha}$. 
}
We will refer to $\Phi$ as the dilaton for reasons that will become clear shortly.

Inserting the components of the chiral superfield $\boldsymbol{\lambda}_\alpha$ in \eqref{E:gotlambdatheta} and \eqref{E:gotlambdathetaab} and the components of the vierbein multiplet given in \eqref{E:FZcomponents} into the expression for the interaction Lagrangian \eqref{E:SKintwithlambda} and integrating over superspace we have
\begin{multline}
\label{E:LintSK}
	\mathcal{L}_{int} = -\frac{1}{4} \varepsilon_{\mu }{}^{\sigma \nu\rho}\delta{}^{a}{}_{\nu} \delta{}^{b}{}_{\rho} \omega_{\sigma ab}J^{\mu} + h{}^{a}{}_{\mu}\delta_{a}{}^{\rho}\eta_{\rho\nu} \left(T_s^{\mu\nu} + \frac{1}{4}\varepsilon^{\mu\nu\tau\sigma}\partial_{\tau}J_{\sigma}\right)\\
	 - \frac{i}{4} \left(\bar{X}_1 \bar{\Lambda} -  X_1 {\Lambda} \right) 
	-\frac{1}{4}\left(\varepsilon^{\mu\nu\sigma\rho}B_{\sigma\rho}\right)F_{\mu\nu}+\frac{1}{24}\Phi\left(-4 T_{s\nu}{}^{\nu}+6Z\right)\,.
\end{multline}
Since the coupling of the vierbein and spin connection to the matter fields are identical to those in the FZ multiplet, see \eqref{E:LintFZ}, the spin current and stress tensor are identical to those in \eqref{E:SFZ} and \eqref{E:TFZ}. 
That is, the spin current is given by the Hodge dual of the bottom component of the KS stress tensor multiplet. Note that the two form $F_{\mu\nu}$ that appears in the components of the KS multiplet in \eqref{E:SKcomponents} does not contribute to the antisymmetric component of the stress tensor or to the spin current.

As before we can check the consistency of the structure of our multiplet by comparing the equations of motion for the vierbein multiplet to linearized supergravity in the first order formalism. In the previous section we found that the kinetic term for the FZ multiplet is given by \eqref{E:kineticFZ}. Since the constraint we use on $\mathbf{L}_{\alpha}$ given in \eqref{E:SKconstraintFZ} is identical to that of the FZ multiplet, we may use \eqref{E:kineticFZ} as a kinetic term for the vierbein. Thus, a kinetic term for both $\mathbf{E}_{\mu}$ and ${\boldsymbol{\lambda}}_{\alpha}$ which is invariant under the vierbein transformational law \eqref{E:coordinateT} and the associated transformation of the chiral superfield $\boldsymbol{\lambda}_\alpha$ \eqref{E:SKgauge} is
\begin{equation}
	\mathcal{L}_{kin}  =\int d^{4}\theta \mathbf{E}_{\lambda}\mathbf{N}_{FZ}^{\lambda} -\frac{C}{48}\intop d^{4}\theta\left(D^{\alpha}{\boldsymbol{\lambda}}_{\alpha}+\bar{D}_{\dot{\alpha}}{\boldsymbol{\lambda}}^{\dot{\alpha}} -\left[\bar{D},\sigma^{\mu}D\right]\mathbf{E}_{\mu}\right)^{2}  \,.
\end{equation}
A somewhat tedious computation involving integration over superspace, establishes
\begin{align}
\begin{split}
\label{E:LkinSK}
	\mathcal{L}_{kin}&=-\delta_{a}{}^{\rho}h^{a}{}_{\mu}\text{Ein}\left[h\right]^{\mu}{}_{\rho}+\left(\omega_{[\nu\sigma\rho]}-\mathring{\omega}_{[\nu\sigma\rho]}\right)^{2}-C\left(3\partial_{[\nu}B_{\sigma\rho]}-\omega_{[\nu\sigma\rho]}+\mathring{\omega}_{[\nu\sigma\rho]}\right)^{2}\\&+\frac{C}{6}\Phi\left(\partial^{2}\Phi+2\text{Ein}\left[h\right]^{\mu}{}_{\mu}\right)+\frac{2}{3}\left(1-C\right)\Lambda\bar{\Lambda}\,.
\end{split}
\end{align}
where $\hbox{Ein}\left[h\right]^{\mu}{}_{\rho}$ are the linearized Einstein equations given in \eqref{E:linEin}.

When $C=0$ the equations of motion which follow from \eqref{E:LkinSK} reduce to those derived from the bosonic Lagrangian obtained from the FZ multiplet in \eqref{E:Lkin}, hence the $B$-field and dilaton are non dynamical.
For $C \neq 0$, the equations of motion for the vierbein, spin connection, dilaton and $B$-field in the absence of matter and for $C \neq 1$ are
\begin{align}
\begin{split}
\label{E:SKEOMCn1}
    \hbox{Ein}\left[\tilde{h}\right]{}^{\mu\nu}&=0\, ,\\
	\partial_{\rho}\partial^{[\rho}B^{\mu\nu]}&=0\, ,\\
	\partial^{2}\Phi&=0\, , \\
 	\left(\omega^{[\mu\nu\rho]}-\mathring{\omega}^{[\mu\nu\rho]}\right)&=-\frac{3C}{\left(1-C\right)}\partial^{[\mu}B^{\nu\rho]}\,,
\end{split}
\end{align}
where $\tilde{h}^{\mu\nu} = h^{\mu\nu} - \frac{C}{6}\eta^{\mu\nu}\Phi$. That is, after shifting the vierbein we get a set of decoupled equations describing a free (tilde'd) vierbein, a dilaton and a $B$-field, and contorsion which is sourced by the $B$-field. For the somewhat unusual $C=1$ case, we find
\begin{align}
\begin{split}
\label{E:SKEOMC1}
	\hbox{Ein}\left[\tilde{h}\right]{}^{\mu\nu}&=0 \, ,\\
	\partial^{[\mu}B^{\nu\rho]}&=0\, ,\\
	\partial_{\rho}\left(\omega^{[\rho\mu\nu]}-\mathring{\omega}^{[\rho\mu\nu]}\right)&=0\,.
\end{split}	
\end{align}
Now the dilaton degree of freedom has disappeared, the $B$ field is flat and the contorsion tensor is no longer sourced.

To better understand the structure of the equations of motion and the $C=1$ configuration, we note that the Lagrangian \eqref{E:LkinSK} is the linearized version of
\begin{multline}
\label{E:LkinSKStr}
	\mathcal{L}_{kin} =  -|e| \bigg( e^{-\frac{1}{3}C\Phi} \left(R(\omega) + \frac{C}{6}(\partial \Phi)^2 \right)-\frac{2}{3}\left(1-C\right)e^{-\frac{2}{3}C\Phi}\Lambda\bar\Lambda \\ + C\left(3\partial_{[\mu} B_{\nu\rho]}- \omega_{[\mu\nu\rho]} + \mathring{\omega}_{[\mu \nu \rho]}  \right)^2 \bigg) 
	+ \mathcal{O}(\mathbf{E}^3) \,,
\end{multline}
where $\omega_{\mu\nu\rho}$ and $\mathring{\omega}_{\mu\nu\rho}$ are given by the non linear version of \eqref{E:defhmunu}, i.e., $\omega_{\mu\nu\rho}=e^a{}_\nu e^b{}_\rho\omega_{\mu ab}$ etc. In the Einstein frame \eqref{E:LkinSKStr} takes the form
\begin{multline}
\label{E:LkinSKEin}
	\mathcal{L}_{kin} =  -|e| \bigg(R(\omega) -\frac{2}{3}\left(1-C\right)\Lambda\bar\Lambda+ \frac{C\left(1-C\right)}{6} \left(\partial\Phi\right)^2 \\+ C e^{-\frac{1}{3}C\Phi }\left(3\partial_{[\mu} B_{\nu\rho]}- e^{\frac{1}{3}C\Phi }\left(\omega_{[\mu\nu\rho]} - \mathring{\omega}_{[\mu \nu \rho]}  \right)\right)^2 \bigg)+ \mathcal{O}(\mathbf{E}^3) \,.
\end{multline}
Thus, the kinetic term for the dilaton vanishes for $C=1$ and it is no longer a dynamical degree of freedom, at least to the order we are working in; the dependence of the action on the exponentiated Dilaton is speculative and may be corrected by the $\mathcal{O}(\mathbf{E}^3)$ terms. To determine the full action, or a possible coupling between the Dilaton and the $B_{\mu\nu}$ field, we would need to compute the full non linear action.

It is sometimes useful to replace the dependence on $\omega_{\mu}{}^{ab}$ with a dependence on the contorsion tensor, $K_{\mu\nu\rho}$, given by $K_{\mu \nu\rho}  =  {\omega}_{\mu\nu\rho} - \mathring{\omega}_{\mu\nu\rho}$, c.f., \eqref{E:ringed}. In this language, the interaction Lagrangian \eqref{E:LintSK} becomes
\begin{multline}
\label{E:LintSKK}
	\mathcal{L}_{int} = -\frac{1}{4} \varepsilon^{\mu\sigma \nu\rho} K_{\sigma\nu\rho} J_{\mu} + h{}^{a}{}_{\mu}\delta_{a}{}^{\rho}\eta_{\rho\nu} T_s^{\mu\nu} \\
	 - \frac{i}{4} \left(\bar{X}_1 \bar{\Lambda} -  X_1 {\Lambda} \right) 
	-\frac{1}{4}\left(\varepsilon^{\mu\nu\sigma\rho}B_{\sigma\rho}\right)F_{\mu\nu}+\frac{1}{24}\Phi\left(-4 T_{s\nu}{}^{\nu}+6Z\right)\,.
\end{multline}
after integrating by parts and omitting boundary terms. Likewise, the kinetic Lagrangian \eqref{E:LkinSK} becomes
\begin{align}
\begin{split}
\label{E:LkinSKK}
	\mathcal{L}_{kin}&=-\delta_{a}{}^{\rho}h^{a}{}_{\mu}\text{Ein}\left[h\right]^{\mu}{}_{\rho}+K_{[\nu\sigma\rho]}K^{[\nu\sigma\rho]}-C\left(3\partial_{[\nu}B_{\sigma\rho]}-K_{[\nu\sigma\rho]}\right)^{2}\\&+\frac{C}{6}\Phi\left(\partial^{2}\Phi+2\text{Ein}\left[h\right]^{\mu}{}_{\mu}\right)+\frac{2}{3}\left(1-C\right)\Lambda\bar{\Lambda}\,.
\end{split}
\end{align}
The coupling of the $B$-field to the contorsion tensor, $(3\partial_{[\mu} B_{\nu\rho]} - K_{[\mu\nu\rho]})^2$ and the coupling of the contorsion tensor to $J^{\mu}$, $\varepsilon^{\mu\sigma\nu\rho}K_{\sigma\nu\rho}J_{\mu}$ can be thought of as a  coupling to torsion required by supersymmetry.

To further our understanding of the $C=1$ equations of motion it is useful to reproduce the interaction \eqref{E:LintSKK} and kinetic Lagrangian \eqref{E:LkinSKK}  by replacing the chiral compensator field $\boldsymbol{\lambda}_{\alpha}$ with the chiral field $\boldsymbol{\ell}$ such that
\begin{equation}
\label{E:SKintwithell}
	\mathcal{L}_{int} = -\frac{1}{8} \int d^2\theta d^2\bar{\theta} \mathbf{S}_{\mu} \mathbf{E}^{\mu} - \frac{1}{16} \left(\int d^2\theta {\boldsymbol{\ell}} \mathbf{X} + h.c.\right)\,.
\end{equation}
Now \eqref{E:SKintwithell} can be made invariant under the vierbein transformation law \eqref{E:coordinateT} such that only the second constraint \eqref{E:SKconstraintR} is satisfied and the constraint \eqref{E:SKconstraintFZ} is replaced by
\begin{equation}
\label{E:elltransform}
	\delta_L \bar{\boldsymbol{\ell}} = \frac{1}{4} D^2 \bar{D}_{\dot{\alpha}} \bar{L}^{\dot{\alpha}}\,.
\end{equation}

The superfield constraint \eqref{E:SKconstraintR} implies the constraints \eqref{E:Rconstraintcomponents} on the component fields. Thus, 
\begin{align}
\begin{split}
\label{E:Rvariation}
	\delta_a{}^{\mu} \delta_L E_{1\mu} &=  \zeta_{a}\, ,\\
	\delta_a{}^{\mu} \delta_L E_{\theta \bar{\theta}}{}_{\mu b} &= 2\delta_{[a}{}^{\mu} \delta_{b]}{}^{\nu} \partial_{\mu} w_{\nu}  + \delta_b{}^{\mu} \partial_{\mu}\xi_{a} \, , \\
	\delta_L E_{\theta^2 \bar{\theta}^2\mu} &= 4 \partial_{\mu} \hbox{Re} \bar{\theA}_{\dot{\alpha}}{}^{\dot{\alpha}} + 4 \delta^a{}_{\mu} \partial^2 \zeta_{a}  \, ,\\
	\delta_L E_{\bar{\theta}^{2}\mu} &= \bar{\theC}_{a}\delta^a{}_{\mu} \,,
\end{split}
\end{align}
and we may no longer interpret $E_{\theta\bar\theta\,\mu}{}^{b}$ as the vierbein. Instead, we have
\begin{align}
\begin{split}
\label{E:EexpansionR}
	\delta^b{}_{\nu} E_{\theta\bar{\theta}\mu b} & =h_{b(\mu} \delta^b{}_{\nu)} +B_{\mu\nu} \, , \\
	E_{\theta^{2}\bar{\theta}^{2}\mu} & =-2k_{\mu}+4\partial^{2}E_{1\mu} \, ,
\end{split}
\end{align}
where
\begin{align}
\begin{split}
\label{E:deltaLR}
	\delta_L k^{\mu} & =-2\partial^{\mu}\text{Re}{\bar{\theA}}_{\dot{\alpha}}{}^{\dot{\alpha}} \, , \\
	\delta_L B_{\mu\nu} & =2\partial_{[\mu}w_{\nu]}-\partial_{[\mu}\xi_{\nu]}\, .\\
\end{split}
\end{align}
We note that the resulting vierbein multiplet $\mathbf{E}_{\mu}$ differs from the one in \eqref{E:FZcomponents}. Nevertheless, in order to avoid clutter, we use the same notation for both vierbein multiplets and for future ones.

The transformation law \eqref{E:elltransform} reads
\begin{equation}
	\delta_L \ell_1 = \delta_a{}^{\mu} \partial_{\mu}\xi^{a} + i \left( \text{Re} \bar{\theA}_{\dot{\alpha}}{}^{\dot{\alpha}} + 2 \delta_a^{\lambda} \partial_{\lambda}\zeta^a\right) \, ,\qquad
	\delta_L \ell_{\theta^2} = 4i\partial_\mu\theC^\mu\,.
\end{equation}
Thus, we write 
\begin{equation}
\label{E:ell1}
	\ell_1 = h_a{}^{\mu} \delta^a{}_{\mu} + \frac{1}{2}{\Phi} + \frac{i}{2} \left(\rho+4 \partial_{\lambda}E_1^{\lambda}\right) \, ,
\end{equation}
where the Stuckelberg field $\rho$ is added to capture the gauge parameter $\text{Re}\bar{\theA}_{\dot{\alpha}}{}^{\dot{\alpha}}$ and both $\rho$ and the dilaton $\Phi$ are added so that ${\boldsymbol{\ell}}$ is chiral
\begin{equation}
	\delta_L {\Phi} = 0 \, ,
	\qquad
	\delta_L \rho =2\text{Re}\bar{\theA}_{\dot{\alpha}}{}^{\dot{\alpha}}\,.
\end{equation}
Similarly,
\begin{equation}
\label{E:ell2}
	\ell_{\theta^2} = 4 i \left(\Lambda + \partial^{\mu}E_{\theta^2\,\mu}\right) \, ,
\end{equation}
where $\Lambda$ is added so that $\boldsymbol{\ell}$ is chiral
\begin{equation}
	\delta_L \Lambda=0 \, .
\end{equation}
Inserting the components of the vierbein multiplet \eqref{E:EexpansionR}, and the components of the chiral field $\boldsymbol{\ell}$ given in \eqref{E:ell1} and \eqref{E:ell2} into the interaction Lagrangian \eqref{E:SKintwithell} we find
\begin{equation}
\label{E:LintSKR}
	\mathcal{L}_{int} =\frac{1}{4}\left(k_{\mu}+\partial_{\mu}\rho - \varepsilon_{\mu\nu\rho\sigma}\partial^{\nu}B^{\rho\sigma} \right)J^{\mu}+h^{a}{}_\nu \delta_a{}^{\rho} T_{s\nu\rho}+\frac{1}{4}\Phi Z-\frac{1}{4}\varepsilon_{\mu\nu\rho\sigma}B^{\mu\nu}F^{\rho\sigma}-\frac{i}{4}\left(\Lambda X_{1}-h.c.\right) \, .
\end{equation}
Comparing the interaction Lagrangian as computed with the chiral field $\boldsymbol{\ell}$ \eqref{E:LintSKR} to that computed with the chiral field $\boldsymbol{\lambda}_\alpha$ \eqref{E:LintSKK} we find that we should identify
\begin{align}
\begin{split}
\label{E:ktoK}
	k_{\mu}+\partial_{\mu}\rho& = \varepsilon_{\mu\nu\rho\sigma}\left(-K^{\nu\rho\sigma}+\partial^{\nu} B^{\rho\sigma}\right) \, ,\\
\end{split}
\end{align}
and shift the vierbein such that $\delta_a{}^{\nu} h^a{}_{\mu} \to \delta_a{}^{\nu} h^a{}_{\mu} - \frac{1}{3} \delta^{\nu}{}_{\mu} \Phi$\,.

To obtain the kinetic term for the vierbein, we follow the same procedure as before
to obtain
\begin{equation}
\label{E:LkinSKR}
	\mathcal{L}_{kin} =\intop d^{4}\theta \mathbf{E}_{\lambda}\mathbf{N}^{R\lambda}+\frac{1-C}{48}\intop d^{4}\theta\left(\boldsymbol{\ell}+\bar{\boldsymbol{\ell}} -\left[\bar{D},\bar{\sigma}_{\lambda}D\right]\mathbf{E}^{\lambda}\right)^{2} \, ,
\end{equation}
where
\begin{equation}
	\mathbf{N}^{R\lambda}  =\frac{1}{8}\varepsilon^{\lambda\sigma\nu\rho}\bar{\sigma}_{\nu}^{\dot{\delta}\delta}\left[\bar{D}_{\dot{\delta}},D_{\delta}\right]\partial_{\sigma}\mathbf{E}_{\rho} \, ,\\
\end{equation}
which leads to
\begin{align}
\begin{split}
	\mathcal{L}_{kin}   =& -\delta_a{}^{\rho} h^{a}{}_{\mu} \hbox{Ein}\left[h\right]^{\mu}{}_{\rho} + K_{[\nu\rho\sigma]} K^{[\nu\rho\sigma]} - C\left(3\partial^{[\nu}B^{\rho\sigma]}-K^{[\nu\rho\sigma]}\right)^{2}
	\\&
	+\frac{1}{6}\left(1-C\right){\Phi}\left(\partial^{2}\Phi-2\text{Ein}\left[h\right]^{\mu}{}_{\mu}\right)
	+\frac{2}{3}\left(1-C\right)\Lambda\bar\Lambda\,.
\end{split}
\end{align}
Taking $\delta_a{}^{\nu} h^a{}_{\mu} \to \delta_a{}^{\nu} h^a{}_{\mu} - \frac{1}{3} \delta^{\nu}{}_{\mu} \Phi$ reproduces the kinetic Lagrangian computed with the chiral field $\boldsymbol{\lambda}_\alpha$ given by \eqref{E:LkinSKK}.
We can now interpret the $C=1$ configuration, and contrast it with $C=0$. When $C=0$ we have a gravitational action which is invariant under the first constraint \eqref{E:SKconstraintFZ} but not the second constraint \eqref{E:SKconstraintR}, reproducing the FZ kinetic term \eqref{E:Lkin}. When $C=1$ we obtain a gravitational action which is invariant under the second constraint \eqref{E:SKconstraintR} but not the first constraint \eqref{E:SKconstraintFZ}. As we shall see in the next section this corresponds to a kinetic term for the $R$ multiplet.

\subsection{The $R$ multiplet}
Next we consider the $R$ multiplet which can be obtained from the SK multiplet after setting $\mathbf{X}=0$ or, equivalently, $X_1=0$, $Z=0$ and $\partial_{\mu}J^{\mu}=0$. Coupling $\mathbf{R}^{\mu}$ to $\mathbf{E}_{\mu}$ results in an interaction term of the form
\begin{equation}
\label{E:Rcoupling}
	\mathcal{L}_{int} = -\frac{1}{8} \int d^4\theta \mathbf{R}^{\mu} \mathbf{E}_{\mu} \, ,
\end{equation}
similar to the general form of the FZ multiplet \eqref{E:FZcoupling} and KS mulitplet \eqref{E:SKcoupling}. The Lagrangian density \eqref{E:Rcoupling} should be invariant under the transformation law \eqref{E:coordinateT} implying the constraint \eqref{E:SKconstraintR}. As we've seen in the previous section the constraint \eqref{E:SKconstraintR} implies the constraints on the components \eqref{E:Rconstraintcomponents} leading to the transformation of the components of the vierbein given in \eqref{E:Rvariation} and the identification of the components of the vierbein multiplet  \eqref{E:EexpansionR} (with \eqref{E:deltaLR}). Starting from the interaction Lagrangian \eqref{E:LintSKR} and setting the compensator fields $\rho$, $\Phi$ and $\Lambda$ to vanish we find
\begin{equation}
\label{E:LintR}
	\mathcal{L}_{int} = \frac{1}{4} \left(k_{\mu} - \varepsilon_{\mu\nu\rho\sigma}\partial^{\nu}B^{\rho\sigma}\right) J^{\mu} + h^a{}_{\nu} \delta_a{}^{\rho} \tensor{T}{_s^\nu_\rho} -\frac{1}{4}\varepsilon_{\mu\nu\rho\sigma}B^{\mu\nu}F^{\rho\sigma}\,.
\end{equation}
The absence of $\Lambda$ is compatible with the vanishing of $X_1$, the absence of $\Phi$ is compatible with $Z=0$ and the absence of $\rho$ is compatible with $\partial_{\mu}J^{\mu}=0$ after integration by parts. If we now use 
\begin{equation}
\label{E:ktoK2}
	k_{\mu} = \varepsilon_{\mu\nu\rho\sigma}\left(-K^{\nu\rho\sigma}+\partial^{\nu} B^{\rho\sigma}\right)\,,
\end{equation}
inline with \eqref{E:ktoK}, we find that due to the conservation law of $J_{\mu}$ the theory is invariant under the somewhat unusual transformation law for the contorsion tensor $K_{\mu\nu\rho} \to K_{\mu\nu\rho} + \varepsilon_{\mu\nu\rho}{}^{\sigma}\partial_{\sigma}\alpha$.

The kinetic term for the vierbein multiplet is given by the kinetic term obtained in the previous section \eqref{E:LkinSKR} with $C=1$ which follows from the absence of the compensator field $\boldsymbol{\ell}$. Computing the superspace integrals we find
\begin{align}
\begin{split}
\label{E:LkinR}
	\mathcal{L}_{kin}  & = 
	\int d^4\theta \mathbf{E}_{\mu} \mathbf{N}^{R\,\mu}  \, ,
	\\
	&= -\delta_a{}^{\rho} h^{a}{}_{\mu} \hbox{Ein}\left[h\right]^{\mu}{}_{\rho} - 3 \partial^{[\nu}B^{\rho\sigma]}\partial_{[\nu}B_{\rho\sigma]} + k^{\mu} \partial^{\nu}B^{\rho\sigma} \varepsilon_{\mu\nu\rho\sigma} \, , \\
	& = -\delta_a{}^{\rho} h^{a}{}_{\mu} \hbox{Ein}\left[h\right]^{\mu}{}_{\rho} -9  \partial^{[\nu}B^{\rho\sigma]} \partial_{[\nu}B_{\rho\sigma]} +6  \partial_{[\nu}B_{\rho\sigma]}K^{[\nu\rho\sigma]}\,,
\end{split}
\end{align}
where we have used \eqref{E:ktoK2}. 
The second line of \eqref{E:LkinR} is compatible with the linearized new minimal supergravity action \cite{SOHNIUS1981353} 
where the coupling between the vector field $k_{\mu}$ and the $B$ field may be thought of as a mixed Chern-Simons term. 
As expected, when interpreting $k_{\mu}$ as contorsion, c.f., \eqref{E:ktoK2}, $\mathcal{L}_{kin}$ is invariant under $K_{\mu\nu\rho} \to K_{\mu\nu\rho} + \varepsilon_{\mu\nu\rho}{}^{\sigma}\partial_{\sigma}\alpha$.

\subsection{The conformal multiplet}
Finally, we consider the conformal case where both $\mathbf{X}=0$ and ${\boldsymbol{\chi}}_{\alpha} = 0$. In this case we denote the stress tensor multiplet by $\mathbf{C}^{\mu}$. Often, the structure of this multiplet is characterized as a representation of the full superconformal group, see, e.g., \cite{Cordova:2016emh}. For the present purpose, following \eqref{E:FZcoupling} or \eqref{E:Rcoupling}, consider
\begin{equation}
\label{E:Ccoupling}
	\mathcal{L}_{int} = -\frac{1}{8} \int d^4\theta \mathbf{C}^{\mu} \mathbf{E}_{\mu}\,.
\end{equation}
The Lagrangian density \eqref{E:Ccoupling} is invariant under the transformation law \eqref{E:coordinateT} without imposing further restrictions on $\mathbf{L}_{\alpha}$. Thus, using the components \eqref{E:Ldef} of the superfield $\boldsymbol{L}_\alpha$, we find
\begin{align}
\begin{split}
	\delta^b{}_{\nu}E_{\theta\bar{\theta}\mu\,b} &= \delta_b{}^{\mu}h^{a}{}_{\mu} \, ,\\
	E_{\theta^2 \bar{\theta}^2\mu} & = 2 \varepsilon_\mu{}^{\lambda\nu\rho}\delta^a{}_\nu\delta^b{}_\rho\omega_{\lambda ab} + 4 \partial^2 E_{1\mu}\,,
\end{split}
\end{align}
where now $\phi \equiv \hbox{Im}\bar{\theA}_{\dot{\alpha}}{}^{\dot{\alpha}}$ 
may be thought of as a Weyl rescaling. Indeed, under a Weyl rescaling we have $\delta_W e^a{}_{\mu} = \phi e^a{}_{\mu}$ and $\delta_W \omega_{\mu}{}^{ab} = 2 e^{[a}{}_{\mu} e^{b] \nu} \partial_{\nu}\phi$ (see \cite{Buchbinder:1985ux,Shapiro:2001rz,Gallegos:2022jow}) which is compatible with the transformations of the components given in \eqref{deltaE}.

Integrating \eqref{E:Ccoupling} over superspace, the interaction Lagrangian takes the form
\begin{equation}
	\mathcal{L}_{int} = -\frac{1}{4} \varepsilon_{\mu }{}^{\sigma \nu\rho}\delta{}^{a}{}_{\nu} \delta{}^{b}{}_{\rho} \omega_{\sigma ab}J^{\mu} + h{}^{a}{}_{\mu}\delta_{a}{}^{\rho}\eta_{\rho\nu} \left(T_s^{\mu\nu} + \frac{1}{4}\varepsilon^{\mu\nu\tau\sigma}\partial_{\tau}J_{\sigma}\right) \,.
\end{equation}
This interaction term is compatible both with a $B_{\mu\nu} \to 0$ limit of the R multiplet interaction Lagrangian \eqref{E:LintR} (using the definition \eqref{E:ktoK2}) and an $X_1 \to 0$ limit of the FZ multiplet interaction Lagrangian \eqref{E:LintFZ}. 
Since conformal gravity is quartic in the gravitational fields we do not attempt to study the linearized gravitational equations of motion here.

\section{Summary and discussion}
\label{S:discussion}

In this work we have found a consistent method for computing a spin current, $S^{\mu\nu\rho}$, in a supersymmetric setting. In particular, we have argued that the bottom component of the Komargodski-Seiberg (KS) stress tensor multiplet can be considered as the Hodge dual of the spin current. 
In \cite{Dumitrescu:2011iu} a  more general stress tensor multiplet was proposed which subsumes the SK multiplet and, presumably, is the most general one containing $J_Q^{\mu}$, $T^{\mu\nu}$ and no operators of higher spin, and is indecomposable. The multiplet of \cite{Dumitrescu:2011iu} can always be reduced, locally, to the KS multiplet: being somewhat concise, the multiplet of \cite{Dumitrescu:2011iu} reduces to the KS multiplet if a certain $\mathbf{Y}_{\alpha}$ can be written as $\mathbf{Y}_{\alpha} = D_{\alpha} \mathbf{X}$ with $\mathbf{X}$ defined in \eqref{E:SKdef}. Thus, we expect that, at least the coupling of the stress tensor multiplet to the vierbein multiplet will not be significantly modified. 

Indeed, this work focused on identifying the coupling between the spin connection, which sits in the vierbein multiplet, to the spin current, which sits in the stress tensor multiplet. Knowledge of the former leads to the latter. The spin current we obtain by this method is completely antisymmetric. As discussed in the main text, one may always add an improvement term to the spin current, compatible with the supersymmetry algebra, which may contribute to the non antisymmetric components of $S^{\mu\nu\rho}$. The virtue of the spin current we propose is that it is compatible with a supersymmetric coupling of the stress tensor to background torsion.
An alternate method for identifying the spin current would be to focus on the conserved current associated with rotational invariance while still maintaining supersymmetry, similar to the work carried out in \cite{Magro:2001aj}.

In \cite{Magro:2001aj} it was shown that invariance of the action under rotations leads to a conservation equation whose structure is similar to \eqref{E:flatWardprime}. Combining this equation with the conservation equation associated with translations, the authors of \cite{Magro:2001aj} were able to construct a decomposable stress tensor multiplet, which reduces to the one in \cite{Dumitrescu:2011iu}. The virtue of using (a supersymmetric extension of) Noether's method to identify the spin current is that this method can be applied without leaning on the equations of motion. Thus, it will allow us to distinguish between the manifestly symmetric stress tensor $T_{s}^{\mu\nu}$ and what we referred to as $T^{\prime\prime\mu\nu}$ in \eqref{E:flatWardprimeprime}.

So far, our identification of a spin current as the bottom component of the stress tensor multiplet applies to $\mathcal{N}=1$ supersymmetric theories in $3+1$ spacetime dimensions. We have considered $\mathcal{N}=1$, $d=4$ theories since they are ``closest'' to the strong and electroweak theory which controls the dynamics of quarks in heavy ion collisions. It would, however, be interesting, to extend this formalism to other dimensions or (and) theories with more supersymmetry.

Within the context of holography, it would be particularly interesting to understand the structure of the spin current in $\mathcal{N}=2$ or, better yet, $\mathcal{N}=4$ superconformal theories in $4$ dimensions. Extended, $\mathcal{N}=2$ superconformal symmetry possesses an $SU(2)\times U(1)$ $R$-symmetry. One may expect that decomposing the $\mathcal{N}=2$ theory into $\mathcal{N}=1$ multiplets, the spin current will be associated with Hodge dual of the latter $U(1)$ component of the $R$ symmetry. In contrast, $\mathcal{N}=4$ supersymmetric theories posses a $SU(4)$ symmetry. Therefore, if we decompose the $\mathcal{N}=4$ symmetry into $\mathcal{N}=1$ multiplets, any spin current we define using the Hodge dual of the $R$ current will break the $SU(4)$ $R$-symmetry. Since angular momentum should not be charged under the $R$ symmetry, this implies that either the $\mathcal{N}=4$ theory does not posses a natural spin current, or that there is another spin current compatible with the symmetries of the $\mathcal{N}=4$ theory. Since we have no claims about uniqueness of the spin current both possibilities are viable. It would be interesting to, perhaps, use the Noether method of \cite{Magro:2001aj} to evaluate the spin current of the $\mathcal{N}=4$ theory. 
It would be interesting to, perhaps, use the Noether method of \cite{Magro:2001aj} to evaluate the spin current of the $\mathcal{N}=4$ theory, or to carry out an analysis similar to the one in \eqref{E:algebracurrents} to see whether or not $\mathcal{N}=4$ spin currents exist.\footnote{In all likelihood, the $\mathcal{N}=4$ theory probably does not posses a spin current due to the absence of an appropriate dimension 3 conserved current. We thank Z. Komargodski for pointing this out to us.}

Recall that in a holographic context, each operator in the boundary theory is in one-to-one correspondence with a bulk field. The boundary value of the bulk field is associated with the external source conjugate to the operator, and the near boundary behavior of the bulk field is associated with the expectation value of the operator. Thus, we may refer to the bulk field dual to the spin current as bulk torsion. This type of setup for studying torsion in a holographic context was pioneered in \cite{Gallegos:2020otk} (see also \cite{Erdmenger:2022nhz,Erdmenger:2023hne,Aviles:2023igk}). 
The analysis carried out in this work offers a precise mapping between boundary torsion and bulk fields. For definiteness, we focus on dual pairs with $\mathcal{N}=1$, $d=4$ supersymmetry, e.g., the $T^{1,1}$ theory of \cite{Klebanov:1998hh}. The $R$ currents holographic counterpart is an Abelian gauge field $A_{M}$. Thus, the spin current, which is the Hodge dual of the $R$ current can be read off of the Hodge dual of the expectation value of the current dual to $A_{M}$. This relation might hint at an analogy between bulk torsion and $A_M$. Perhaps, with some work, one might be able to extend such an analogy to bulk torsion in supergravity or higher derivative gravity theories. 
In this context it would be interesting to check how ghosts associated with dynamical torsion (as found in, e.g., \cite{Jimnez2020}) are resolved.

Also related to holography is the interplay between a bulk $B$ field and torsion. In this work and many others, e.g., \cite{Howe:1984fak,Curtright:1984dz,Braaten:1985is,Gates:1984nk,Strominger:1986uh,Hull:1987pc,Katanaev:1986qm}, the $B$-field seems to be tied to dynamical torsion. It would be satisfying if one can relate the holographic $B$ field to torsion on the boundary in-line with what we found in \eqref{E:SKEOMCn1}.  If bulk torsion may be extended beyond a holographic description, it may be possible to recast the analysis in this work as a rigid limit of such a theory similar to what was done in \cite{Festuccia:2011ws} to describe supersymmetric theories in a curved background.

Given that one can identify torsion with the Hodge dual of the $R$ current, at least for $\mathcal{N}=1$, $d=4$ theories, one can now proceed to apply this result to various settings including, but not limited to, hydrodynamics of spin currents \cite{Voloshin:2004ha,Becattini:2007sr,Betz:2007kg,Florkowski:2017ruc,Hongo:2021ona,Gallegos:2021bzp,Gallegos:2022jow}. We expect to report on such applications in the near future.

\section*{Acknowledgements}
We would like to thank G. Festuccia, N. Iqbal, Z. Komargodski, and G. Zafrir, for useful discussions. RK is supported in part by the Miriam and Aaron Gutwirth Memorial Fellowship. RK and AY are supported in part by an ISF-NSFC grant 3191/23, and a BSF grant 2022110. CC and UG are supported by the Netherlands Organisation for Scientific Research (NWO) under the VICI grant VI.C.202.104. DG is supported by Universidad Nacional Aut\'onoma de M\'exico through the program ``Programa de Becas Posdoctorales en la UNAM."

\begin{appendix}

\section{SUSY}
\label{A:SUSY}
We use conventions where greek indices starting from $\mu$, viz., $\mu,\nu,\ldots$ specify coordinates on the (flat) spacetime manifold and lower case roman letters $a,b,\ldots$ specify tangent space indices. Components of right handed  Weyl spinors are denoted by greek indices, starting from $\alpha$, i.e., $\alpha,\beta$ and those of left handed Weyl spinors are dotted versions of the latter. 

For any Grassmanian quantity $\psi_{\alpha}$ or ${\bar{\psi}}_{\dot\alpha}$ we define:
\begin{align}
\begin{split}
	\psi^{\alpha}=\varepsilon^{\alpha\beta}\psi_{\beta}\, ,&\qquad\psi_{\alpha}=\varepsilon_{\alpha\beta}\psi^{\beta}\,,\\
    {\bar\psi}^{\dot\alpha}=\varepsilon^{\dot\alpha\dot\beta}{\bar\psi}_{\dot\beta}\, ,&\qquad{\bar\psi}_{\dot\alpha}=\varepsilon_{\dot\alpha\dot\beta}{\bar\psi}^{\dot\beta}\,, 
\end{split}
\end{align}
where $\varepsilon^{12}=-\varepsilon^{21}=-\varepsilon_{12}=\varepsilon_{21}=1$ and $0$ otherwise. The Pauli matrices $\sigma^a_{\alpha\dot\alpha}$ are given by
\begin{equation} \label{E:Pauli}
    \sigma^{0}=-\begin{pmatrix}
	1 & 0\\
	0 & 1
	\end{pmatrix}
    \,,\qquad
    \sigma^{1}=\begin{pmatrix}
	0 & 1\\
	1 & 0
	\end{pmatrix}
    \,,\qquad
    \sigma^{2}=\begin{pmatrix}
	0 & -i\\
	i & 0
	\end{pmatrix}
    \,,\qquad
    \sigma^{3}=\begin{pmatrix}
	1 & 0\\
	0 & -1
	\end{pmatrix}\,,
\end{equation}
and we also define
\begin{equation}
\bar{\sigma}^{a\dot{\alpha}\alpha}=\varepsilon^{\alpha\beta}\varepsilon^{\dot{\alpha}\dot{\beta}}\sigma_{\beta\dot{\beta}}^{a}\,.
\end{equation}
From the Pauli matrices we construct $\sigma^{ab}$ and $\bar{\sigma}^{ab}$,
\begin{equation}\label{E:sigmaDef}
    \sigma^{ab}=\frac{1}{4}\left(\sigma^{a}\bar{\sigma}^{b}-\sigma^{b}\bar{\sigma}^{a}\right)
    \, ,\qquad
    \bar{\sigma}^{ab}=\frac{1}{4}\left(\bar{\sigma}^{a}\sigma^{b}-\bar{\sigma}^{b}\sigma^{a}\right)\,,
\end{equation}
which allow us to write the covariant derivative of a spinor as
\begin{align}
    \nabla_{\mu}\psi&=\left(\partial_{\mu}-\frac{1}{2}\sigma^{ab}\omega_{\mu ab}\right)\psi\,,\\\nabla_{\mu}\bar{\psi}&=\bar{\psi}\left(\overleftarrow{\partial}_{\mu}+\frac{1}{2}\bar{\sigma}^{ab}\omega_{\mu ab}\right)\,,
\end{align}
with $\bar{\psi}\overleftarrow{\partial}_{\mu}=\partial_{\mu}\bar{\psi}$.

Our conventions for superspace coordinates are given by
\begin{align}
	\frac{\partial}{\partial\theta^{\alpha}}\theta^{\beta}&=\delta_{\alpha}^{\beta}\,,
	&\frac{\partial}{\partial\bar{\theta}^{\dot{\alpha}}}\bar{\theta}^{\dot{\beta}}&=\delta_{\dot{\alpha}}^{\dot{\beta}}\,,\\
	\intop d^{2}\theta\theta^{2}&=\varepsilon^{\beta\alpha}\frac{\partial}{\partial\theta^{\alpha}}\frac{\partial}{\partial\theta^{\beta}}\theta^{2}=-4\,,
	&\intop d^{2}\bar{\theta}\bar{\theta}^{2}&=\varepsilon^{\dot{\alpha}\dot{\beta}}\frac{\partial}{\partial\bar{\theta}^{\dot{\alpha}}}\frac{\partial}{\partial\bar{\theta}^{\dot{\beta}}}\bar{\theta}^{2}=-4\,.
\end{align}
Supersymmetry multiplets are denoted by boldface letters. If $\mathbf{A}$ is a supersymmetric multiplet, we define its components via
\begin{multline}
	\mathbf{A} = A_1 + \theta^{\beta}A_{\theta\,\beta}+\bar{\theta}_{\dot{\beta}} A_{\bar{\theta}}^{\dot{\beta}} + \frac{1}{2} \theta \sigma_{b} \bar{\theta} A_{\theta\bar{\theta}}^{b} - \frac{1}{4} \theta^2 A_{\theta^2}-\frac{1}{4} \bar{\theta}^2 A_{\bar{\theta}^2} \\
	- \frac{1}{4} \bar{\theta}^2 \theta^{\beta} A_{\bar{\theta}^2\theta\,\beta} - \frac{1}{4} \theta^2 \bar{\theta}_{\dot{\beta}} A_{\theta^2\bar{\theta}}^{\dot{\beta}} + \frac{1}{16} \theta^2 \bar{\theta}^2 A_{\theta^2\bar{\theta}^2} \,\, .
\end{multline}
A superderivative acts on a multiplet via 
\begin{align}
\begin{split}
    D&=\frac{\partial}{\partial\theta}+i\sigma^\mu \bar\theta \mathring{\nabla}_\mu\,,\\
    \bar D&=-\frac{\partial}{\partial\bar\theta}+i{\bar\sigma}^\mu \theta \mathring{\nabla}_\mu\,.   
\end{split}
\end{align}

\end{appendix}

\bibliographystyle{JHEP}
\bibliography{SUSYspin_submission}

\end{document}